\documentclass[aps,prl,twocolumn,superscriptaddress,showpacs]{revtex4} 
\newcommand{\figsize}{0.45\textwidth}      
\usepackage{graphicx}
\usepackage{dcolumn}
\usepackage{bm}
\usepackage{setspace}

\newcommand{\ttbar} {\ensuremath{t\bar{t}} }
\newcommand{\ttbars}{\ensuremath{t\bar{t}}}

\newcommand{\ppbar} {\ensuremath{p\bar{p}} }
\newcommand{\ttoHb} {\ensuremath{t\rightarrow H^{+}b} }

\newcommand{\ttoHbs}{\ensuremath{t\rightarrow H^{+}b}}
\newcommand{\ttoWbs}{\ensuremath{t\rightarrow W^{+}b}}

\newcommand{\brttoHb}{BR(\ttoHbs) }
\newcommand{\brttoHbs}{BR(\ttoHbs)}

\newcommand{\tanb} {\ensuremath{\tan\beta} }

\newcommand{\tanbs}{\ensuremath{\tan\beta}}

\newcommand{\Wbbbar} {\ensuremath{W^{+}b\bar{b}} }
\newcommand{\csbar}  {\ensuremath{c\bar{s}} }
\newcommand{\csbars} {\ensuremath{c\bar{s}}}
\newcommand{\taunu}  {\ensuremath{\bar{\tau}\nu} }
\newcommand{\taunus} {\ensuremath{\bar{\tau}\nu}}
\newcommand{\tstarb} {\ensuremath{t^*\bar{b}} }

\newcommand{\Whzero} {\ensuremath{W^+h^0} }
\newcommand{\WAzero} {\ensuremath{W^+A^0} }
\newcommand{\Whzeros}{\ensuremath{W^+h^0}}
\newcommand{\WAzeros}{\ensuremath{W^+A^0}}

\newcommand{\Htotaunu}  {\ensuremath{H^+\rightarrow \taunu} }
\newcommand{\Htotstarb} {\ensuremath{H^+\rightarrow \tstarb} }
\newcommand{\HtoWhzero} {\ensuremath{H^+\rightarrow \Whzero} }

\newcommand{\Htocsbars} {\ensuremath{H^+\rightarrow \csbar}}
\newcommand{\Htotaunus} {\ensuremath{H^+\rightarrow \taunu}}
\newcommand{\Htotstarbs}{\ensuremath{H^+\rightarrow \tstarb}}
\newcommand{\HtoWhzeros}{\ensuremath{H^+\rightarrow \Whzero}}

\newcommand{\Hpl} {\ensuremath{H^{+}} }

\newcommand{\HplHmi} {\ensuremath{H^{+}H^{-}} }
\newcommand{\Hpm} {\ensuremath{H^{\pm}} }
\newcommand{\Hpms}{\ensuremath{H^{\pm}}}
\newcommand{\mH}  {\ensuremath{m_{H^{\pm}}} }
\newcommand{\mHs} {\ensuremath{m_{H^{\pm}}}}
\newcommand{\mt}  {\ensuremath{m_{t}} }

\newcommand{\gevcc}{\ensuremath{\text{GeV/}c^2} }
\newcommand{\gevccs}{\ensuremath{\text{GeV/}c^2}}

\newcommand{\invpb}{\ensuremath{\text{pb}^{-1}} }
\newcommand{\invpbs}{\ensuremath{\text{pb}^{-1}}}

\newcommand{\met}{\ensuremath{E_{T}\hspace{-0.175in}/\hspace{0.1in}} }

\newcommand{\ninefivecl}{95\% C.L. }

\begin{document}


\preprint{CDF/PUB/TOP/CDFR/7778} 
\preprint{Version 3.3}          

\title{Search for charged Higgs bosons from top quark decays in \\ 
\ppbar collisions at $\sqrt{s}=1.96$~TeV}

\date{\today}

\begin{abstract}

We report the results of a search for a charged Higgs boson in the decays of top quarks produced in \ppbar collisions at a center-of-mass energy of 1.96~TeV. We use a data sample corresponding to an integrated luminosity of 193 \invpb collected by the upgraded Collider Detector at Fermilab.
No evidence for charged Higgs production is found, allowing \ninefivecl upper limits to be placed on \brttoHb for different charged Higgs decay scenarios.  
In addition, we present in the MSSM (\mHs,\tanbs) plane the first exclusion regions with radiative and Yukawa coupling corrections.

\end{abstract}




\affiliation{Institute of Physics, Academia Sinica, Taipei, Taiwan 11529, Republic of China} 
\affiliation{Argonne National Laboratory, Argonne, Illinois 60439} 
\affiliation{Institut de Fisica d'Altes Energies, Universitat Autonoma de Barcelona, E-08193, Bellaterra (Barcelona), Spain} 
\affiliation{Baylor University, Waco, Texas  76798} 
\affiliation{Istituto Nazionale di Fisica Nucleare, University of Bologna, I-40127 Bologna, Italy} 
\affiliation{Brandeis University, Waltham, Massachusetts 02254} 
\affiliation{University of California, Davis, Davis, California  95616} 
\affiliation{University of California, Los Angeles, Los Angeles, California  90024} 
\affiliation{University of California, San Diego, La Jolla, California  92093} 
\affiliation{University of California, Santa Barbara, Santa Barbara, California 93106} 
\affiliation{Instituto de Fisica de Cantabria, CSIC-University of Cantabria, 39005 Santander, Spain} 
\affiliation{Carnegie Mellon University, Pittsburgh, PA  15213} 
\affiliation{Enrico Fermi Institute, University of Chicago, Chicago, Illinois 60637} 
\affiliation{Joint Institute for Nuclear Research, RU-141980 Dubna, Russia} 
\affiliation{Duke University, Durham, North Carolina  27708} 
\affiliation{Fermi National Accelerator Laboratory, Batavia, Illinois 60510} 
\affiliation{University of Florida, Gainesville, Florida  32611} 
\affiliation{Laboratori Nazionali di Frascati, Istituto Nazionale di Fisica Nucleare, I-00044 Frascati, Italy} 
\affiliation{University of Geneva, CH-1211 Geneva 4, Switzerland} 
\affiliation{Glasgow University, Glasgow G12 8QQ, United Kingdom} 
\affiliation{Harvard University, Cambridge, Massachusetts 02138} 
\affiliation{Division of High Energy Physics, Department of Physics, University of Helsinki and Helsinki Institute of Physics, FIN-00014, Helsinki, Finland} 
\affiliation{University of Illinois, Urbana, Illinois 61801} 
\affiliation{The Johns Hopkins University, Baltimore, Maryland 21218} 
\affiliation{Institut f\"{u}r Experimentelle Kernphysik, Universit\"{a}t Karlsruhe, 76128 Karlsruhe, Germany} 
\affiliation{High Energy Accelerator Research Organization (KEK), Tsukuba, Ibaraki 305, Japan} 
\affiliation{Center for High Energy Physics: Kyungpook National University, Taegu 702-701; Seoul National University, Seoul 151-742; and SungKyunKwan University, Suwon 440-746; Korea} 
\affiliation{Ernest Orlando Lawrence Berkeley National Laboratory, Berkeley, California 94720} 
\affiliation{University of Liverpool, Liverpool L69 7ZE, United Kingdom} 
\affiliation{University College London, London WC1E 6BT, United Kingdom} 
\affiliation{Massachusetts Institute of Technology, Cambridge, Massachusetts  02139} 
\affiliation{Institute of Particle Physics: McGill University, Montr\'{e}al, Canada H3A~2T8; and University of Toronto, Toronto, Canada M5S~1A7} 
\affiliation{University of Michigan, Ann Arbor, Michigan 48109} 
\affiliation{Michigan State University, East Lansing, Michigan  48824} 
\affiliation{Institution for Theoretical and Experimental Physics, ITEP, Moscow 117259, Russia} 
\affiliation{University of New Mexico, Albuquerque, New Mexico 87131} 
\affiliation{Northwestern University, Evanston, Illinois  60208} 
\affiliation{The Ohio State University, Columbus, Ohio  43210} 
\affiliation{Okayama University, Okayama 700-8530, Japan} 
\affiliation{Osaka City University, Osaka 588, Japan} 
\affiliation{University of Oxford, Oxford OX1 3RH, United Kingdom} 
\affiliation{University of Padova, Istituto Nazionale di Fisica Nucleare, Sezione di Padova-Trento, I-35131 Padova, Italy} 
\affiliation{University of Pennsylvania, Philadelphia, Pennsylvania 19104} 
\affiliation{Istituto Nazionale di Fisica Nucleare Pisa, Universities of Pisa, Siena and Scuola Normale Superiore, I-56127 Pisa, Italy} 
\affiliation{University of Pittsburgh, Pittsburgh, Pennsylvania 15260} 
\affiliation{Purdue University, West Lafayette, Indiana 47907} 
\affiliation{University of Rochester, Rochester, New York 14627} 
\affiliation{The Rockefeller University, New York, New York 10021} 
\affiliation{Istituto Nazionale di Fisica Nucleare, Sezione di Roma 1, University of Rome ``La Sapienza," I-00185 Roma, Italy} 
\affiliation{Rutgers University, Piscataway, New Jersey 08855} 
\affiliation{Texas A\&M University, College Station, Texas 77843} 
\affiliation{Istituto Nazionale di Fisica Nucleare, University of Trieste/\ Udine, Italy} 
\affiliation{University of Tsukuba, Tsukuba, Ibaraki 305, Japan} 
\affiliation{Tufts University, Medford, Massachusetts 02155} 
\affiliation{Waseda University, Tokyo 169, Japan} 
\affiliation{Wayne State University, Detroit, Michigan  48201} 
\affiliation{University of Wisconsin, Madison, Wisconsin 53706} 
\affiliation{Yale University, New Haven, Connecticut 06520} 
\author{A.~Abulencia}
\affiliation{University of Illinois, Urbana, Illinois 61801}
\author{D.~Acosta}
\affiliation{University of Florida, Gainesville, Florida  32611}
\author{J.~Adelman}
\affiliation{Enrico Fermi Institute, University of Chicago, Chicago, Illinois 60637}
\author{T.~Affolder}
\affiliation{University of California, Santa Barbara, Santa Barbara, California 93106}
\author{T.~Akimoto}
\affiliation{University of Tsukuba, Tsukuba, Ibaraki 305, Japan}
\author{M.G.~Albrow}
\affiliation{Fermi National Accelerator Laboratory, Batavia, Illinois 60510}
\author{D.~Ambrose}
\affiliation{Fermi National Accelerator Laboratory, Batavia, Illinois 60510}
\author{S.~Amerio}
\affiliation{University of Padova, Istituto Nazionale di Fisica Nucleare, Sezione di Padova-Trento, I-35131 Padova, Italy}
\author{D.~Amidei}
\affiliation{University of Michigan, Ann Arbor, Michigan 48109}
\author{A.~Anastassov}
\affiliation{Rutgers University, Piscataway, New Jersey 08855}
% ...
\collaboration{CDF Collaboration}
\noaffiliation

\maketitle

\newpage
\enlargethispage{0.5cm}

One of the open questions in the standard model (SM) of particle physics involves the mechanism of electroweak symmetry breaking (EWSB).  
Within the SM it is postulated that a single scalar doublet field breaks the symmetry, resulting in a single observable particle of unknown mass called the Higgs boson \cite{Higgs:1964ia}.
To date, the SM Higgs boson has not been observed, and extensions of the SM have been proposed with different Higgs phenomenologies. 
The simplest extension of the SM Higgs sector introduces another Higgs doublet.
In two-Higgs doublet models  EWSB results in five Higgs bosons, three of which are neutral ($h^0,H^0,A^0$) and two of which are charged (\Hpms). 
The minimal supersymmetric extension of the SM (MSSM) includes a two-Higgs doublet sector in which one doublet couples to the up-type quarks and neutrinos, and the other to the down-type quarks and charged leptons~\cite{MSSM_cite}. 
The observation of a charged Higgs boson would provide unambiguous evidence of a Higgs sector richer than that predicted by the SM. 

At the Tevatron, direct production of single charged Higgs is expected to be negligible, and the direct production of \HplHmi via the weak interaction is  expected to have a relatively small cross section on the order of 0.1~pb~\cite{HBTAP_cite}. 
The production of \ttbar pairs,  with a theoretical production cross section of $6.7^{+0.7}_{-0.9}$~pb~\cite{THEO_SIGMA_cite1,THEO_SIGMA_cite2} for \mt=175~\gevccs, may offer another source of charged Higgs production. If kinematically allowed, the top quark can decay to $H^{+}b$, competing with the SM decay \ttoWbs.  This mechanism can provide a larger production rate of charged Higgs and offers a much cleaner signature than that of direct production.

Previous searches for the charged Higgs boson have been performed at $\sqrt{s}=1.8$ TeV in the $\tau_h+\met+{\rm jets}+\ell$ channels, where the missing energy \met is defined in \cite{MET_cite}, $\tau_{h}$ denotes a tau lepton which decays to hadrons,  and where $\ell=e$ or $\mu$  in \cite{CDF_PRD_cite} and $\ell=e, \mu$ or $\tau_{h}$ in \cite{CDF_PRL_cite}.
In the framework of the tauonic Higgs model, in which the charged Higgs decays exclusively to \taunus, these searches set limits directly on \brttoHb based on the measured production rate. These results are then translated into limits on \tanbs, the ratio of vacuum expectation values of the two Higgs doublets. 
Other searches obtained limits in the (\mHs,\tanbs) plane, assuming that the charged Higgs decays only to \taunu in the  $\met+{\rm jets}+\tau_h$ channel in~\cite{DO_PRL2_cite}, and assuming that the charged Higgs decays to \taunus, \csbars, and \tstarb$(\rightarrow \Wbbbar)$  in the  $\met+{\rm jets}+\ell$ channel with $\ell=e$ or $\mu$ in~\cite{DO_PRL_cite}, where $t^*$ is a virtual top quark. 

These limits utilize tree-level MSSM predictions of the \ttoHb  and  charged Higgs  branching fraction as a function of \tanbs. It is now known that higher-order radiative corrections significantly modify these predictions. The corrections strongly depend on the parameters of the model and are particularly large at high values of \tanb \cite{CARENA_cite}.
In addition, it is also predicted that in the low \tanb region,
the charged Higgs has a sizable branching fraction to \Whzeros.


CDF has recently reported measurements of the \ttbar production cross section in the  $\ell+\met+{\rm jets}+ X$ channels, where $\ell=e, \mu$ and where $X=\ell$ (the ``dilepton'' channel), $X=\tau_{h}$ (``lepton+tau''), $X=$ one or more tagged jets \cite{TAG_cite} (``lepton+jets, $\geq$ 1 tag''), and  $X=$ two or more tagged jets (``lepton+jets, $\geq$ 2 tags''). These measurements are carried out under the assumption \brttoHb = 0 and use data samples corresponding to an integrated luminosity of up to 193 \invpbs~\cite{DILEP_XS_cite, LTAUH_XS_cite, LJETS1_XS_cite}.  Each measurement agrees with the SM \ttbar cross section expectation within its uncertainty, providing no evidence for non-SM physics.
In this analysis, we consider the possibility of \ttoHb and recast the cross section results to set limits on charged Higgs production. Depending on the top and Higgs branching ratios, the number of expected events in these decay channels can show an excess or deficit with respect to SM expectations.

As published, these measurements allow the categorization of a single event in multiple channels. In this analysis extra requirements are applied to each channel in order to force the association of every event to a single channel. The \ttbar signal acceptance and non-\ttbar SM background contribution to each of these exclusive channels are recalculated, and the changes from the original cross section analyses are found to be mostly negligible. The only exception to this is the ``$\geq$ 1 tag'' and ``$\geq$ 2 tags'' lepton+jets channels, where the latter is a proper subset of the former.  Removal of this 100\% overlap changes the ``$\geq$ 1 tag'' channel to exactly one tag,``= 1 tag''. The results for these new exclusive channels in terms of background, number of observed events, and number of SM expected events are shown in Table~\ref{DATA_BACK_label}.

We assume that the charged Higgs boson can decay only  to \taunus, \csbars, \tstarb or \Whzeros, leading to five possible decay modes for a single top quark: (1) \ttoWbs, (2) \ttoHbs, \Htotaunus, (3)  \ttoHbs, \Htocsbars, (4) \ttoHbs, \Htotstarbs, and (5) \ttoHbs, \HtoWhzeros, $\mathrm{h^0\rightarrow b\bar{b}}$. Charge conjugated decays are implied.
\begin{table}
\caption{Number of events in each exclusive channel from non-\ttbar SM background sources, observed in data, and total expected assuming  $\sigma_{\ttbar}^{\rm prod}=6.7$ pb and $\brttoHbs=0$.\label{DATA_BACK_label} }
\begin{center}
\begin{tabular}{c c c c}
\hline
\hline
Channel                       &  Background    &  Data   &  SM-expected   \\
                              &   events       &  events &  events        \\
\hline			    		                                    
dilepton                      &  $2.7\pm0.7$   & 13      & $10.9\pm1.4$   \\
lepton+jets, $=$ 1 tag        &  $21.8\pm3.0$  & 49      & $54.0\pm4.3$   \\
lepton+jets, $\geq$ 2 tags    &  $1.3\pm0.3$   & 8       & $10\pm1$       \\
lepton + tau                  &  $1.3\pm0.2$   & 2       & $2.3\pm0.3$    \\
\hline
\hline
\end{tabular}
\end{center}
\end{table}
Allowing for a non-zero \brttoHbs, the acceptance of the detector for a given \ttbar channel {\it k} is 
\begin{equation}
\mathcal{A}_k = \sum_{i,j=1}^{5} B_i \cdot B_j \cdot 
\epsilon_{ij,k}(\Gamma_{t},\Gamma_{H^{\pm}},m_{H^{\pm}},m_{h^0}),
\end{equation}
where $B_i$ ($B_j$) represent the branching fractions of the top quark (anti-quark) to decay via mode $i$ ($j$) as listed above, and $\epsilon_{ij,k}$ is the efficiency to detect a \ttbar event whose top quarks decay via  modes $i$ and $j$ in channel {\it k}. 


The efficiencies $\epsilon_{ij,k}$ are obtained from Monte Carlo (MC) simulation of \ttbar events  generated with different masses of the top, \Hpms, and $h^0$. The MC generator {\sc PYTHIA} \cite{PYTHIA_cite} is modified to include the decay \Htotstarb and is used for the generation of the \ttbar events.
The detector simulation and reconstruction algorithms for muons, electrons, and jets are identical to those used in the SM \ttbar cross section measurements for the four channels. MC efficiencies are scaled for known differences between MC simulation of the detector response and that observed in data.
The dependence of the efficiencies on the width of the top quark ($\Gamma_{t}$) and the width of the charged Higgs ($\Gamma_{H^{\pm}}$) is taken into account using the simulated \ttbar events.
The systematic uncertainties on $\epsilon_{ij,k}$ for the process $\ttbars\rightarrow W^+b\,W^-\bar{b}$ are listed in \cite{DILEP_XS_cite, LTAUH_XS_cite, LJETS1_XS_cite} and do not differ much for the other possible decay modes.

The expected number of events in channel {\it k} is  
\begin{equation}
  \mu_{k} = \sigma_{\ttbar}^{\rm prod} \cdot \mathcal{A}_{k}(\rho) \cdot \mathcal{L}_{k} + n_{k}^{\rm back},
\end{equation}
where $\sigma_{\ttbar}^{\rm prod}$ is the \ttbar production cross section and  $\rho$ represents a generic model from which the nine quantities (five BR's, $\Gamma_{t}$, $\Gamma_{H^{\pm}}$, \mH and $m_{h^0}$), needed to calculate the acceptance $\mathcal{A}_k$, can be derived.  $\mathcal{L}_{k}$ is the integrated luminosity, and $ n_{k}^{back}$ is the number of expected background events in channel {\it k} (shown in Table~\ref{DATA_BACK_label}).
We assume the inclusion of the Higgs sector does not modify the value of the \ttbar production cross section and set it to $\sigma_{\ttbar}^{\rm prod}=6.7\pm 0.9$~pb.

For each channel a  likelihood is constructed based on the Poisson probability to observe $N_k$ events when a given model predicts $\mu_k$ events.  Since the four channels were constructed to be mutually exclusive, the product of their likelihoods is taken to form a final likelihood. The correlations of the efficiencies, backgrounds, and systematic uncertainties between channels are taken into account. 
The posterior probability distribution of the parameter of interest is constructed from the likelihood and a prior probability density. The posterior probability is integrated to determine the excluded values of the parameter.


In the MSSM the nine quantities needed to calculate the acceptance are predicted from a specific set of MSSM parameters, including $m_{\Hpm}$ and \tanbs. We use the computational package {\sc CPsuperH} \cite{CPSUPERH_cite}  to compute all the Higgs masses and branching ratios. This program includes  QCD, SUSY-QCD, and SUSY-EW radiative corrections up to the two-loop leading logarithms and applies these corrections to the top and bottom Yukawa couplings. The top branching ratio to charged Higgs is computed with the same level of accuracy from custom code developed in collaboration with the authors of \cite{CARENA_cite}.
In the context of the MSSM with $m_{A^0}<\mH$, {\sc CPsuperH} predicts that the \Hpl decay to \WAzero is non-negligible for masses of \Hpm below 100 \gevccs. In this case {\sc CPsuperH} also predicts the mass of the  $A^0$  to be similar to that of the $h^0$, and we assume the kinematics of the decay to \WAzero to be identical to that of \Whzero when the $h^0$ and $A^0$ masses are equal. Thus, we assign to the decay \HtoWhzero a branching ratio of \mbox{BR(\Hpl$\rightarrow$\Whzeros)} + \mbox{BR(\Hpl$\rightarrow$\WAzeros)}, effectively considering both decays.

\begin{figure}
\center
\includegraphics[width=\figsize]{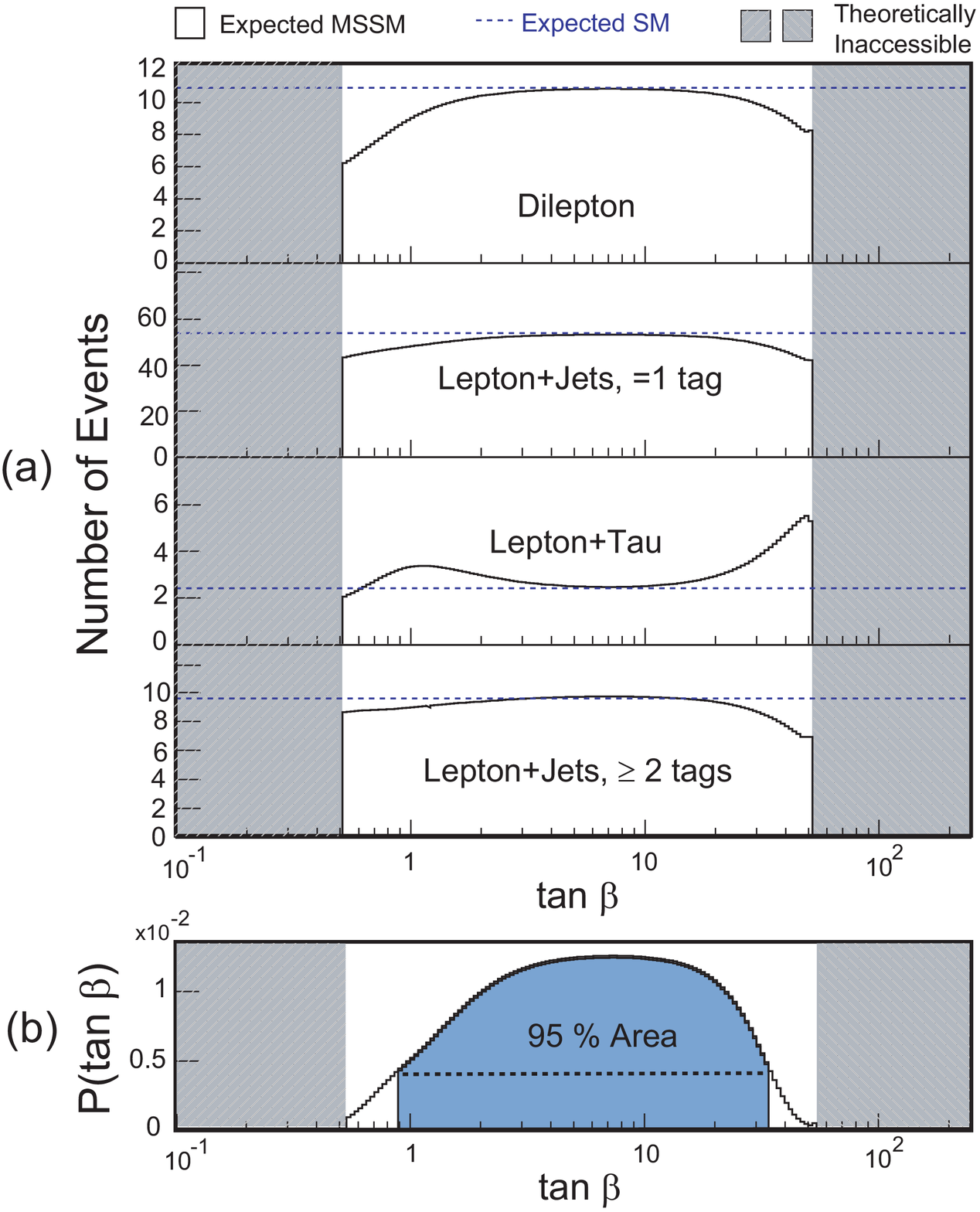}
\caption{ Predictions for $\mH=120$ \gevcc and $\mt=175$ \gevcc as a function of \tanb for 193 \invpbs. The MSSM parameters are defined in~\cite{Gunion:1989we} and are set to $M_{SUSY}=1000$ \gevccs, $\mu=-500$ \gevccs, $A_t=A_b=2000$ \gevccs, $A_\tau=500$ \gevccs, $M_2=M_3=M_Q=M_U=M_D=M_E=M_L=M_{SUSY}$, and $M_1=0.498 M_2$. (a) Expected number of events in each of the channels. (b) Posterior probability density obtained when the number of observed events is equal to that expected from the SM.
\label{ExpectedEvents_fig} }
\end{figure}

As an example of how a charged Higgs alters the balance between the top decay channels, Fig.~\ref{ExpectedEvents_fig}(a) shows the expected number of events in each of the exclusive channels as a function of \tanb for $\mH=120$ \gevccs. The other relevant MSSM parameters are detailed in the caption. 
The figure demonstrates the excess expected in the lepton+tau  channel and the deficit expected in the other channels for large \tanb values. For values of \tanb around 7 the \mbox{BR(\ttoHbs)} goes to zero and the SM expectation for the different channels is recovered. The relationship between the channels changes with charged Higgs mass. 
Values of \tanb which result in a non-self-consistent  Higgs sector are reported by {\sc CPsuperH} and are shown as the theoretically inaccessible regions in Fig.~\ref{ExpectedEvents_fig}.

Figure~\ref{ExpectedEvents_fig}(b) shows the posterior probability obtained for the four channels when the number of observed events is equal to that expected from the SM. The posterior is obtained by means of a flat prior in  $\textrm{log}_{10}(\tanb)$. This prior allows for a smooth variation of the top and charged Higgs branching ratios as a function of $\textrm{log}_{10}(\tanb)$. 
The probability is integrated over its maximum density region to obtain expected upper and lower limits on \tanb  at the 95\% confidence level (C.L.)

\begin{figure}
\center
\includegraphics[width=\figsize]{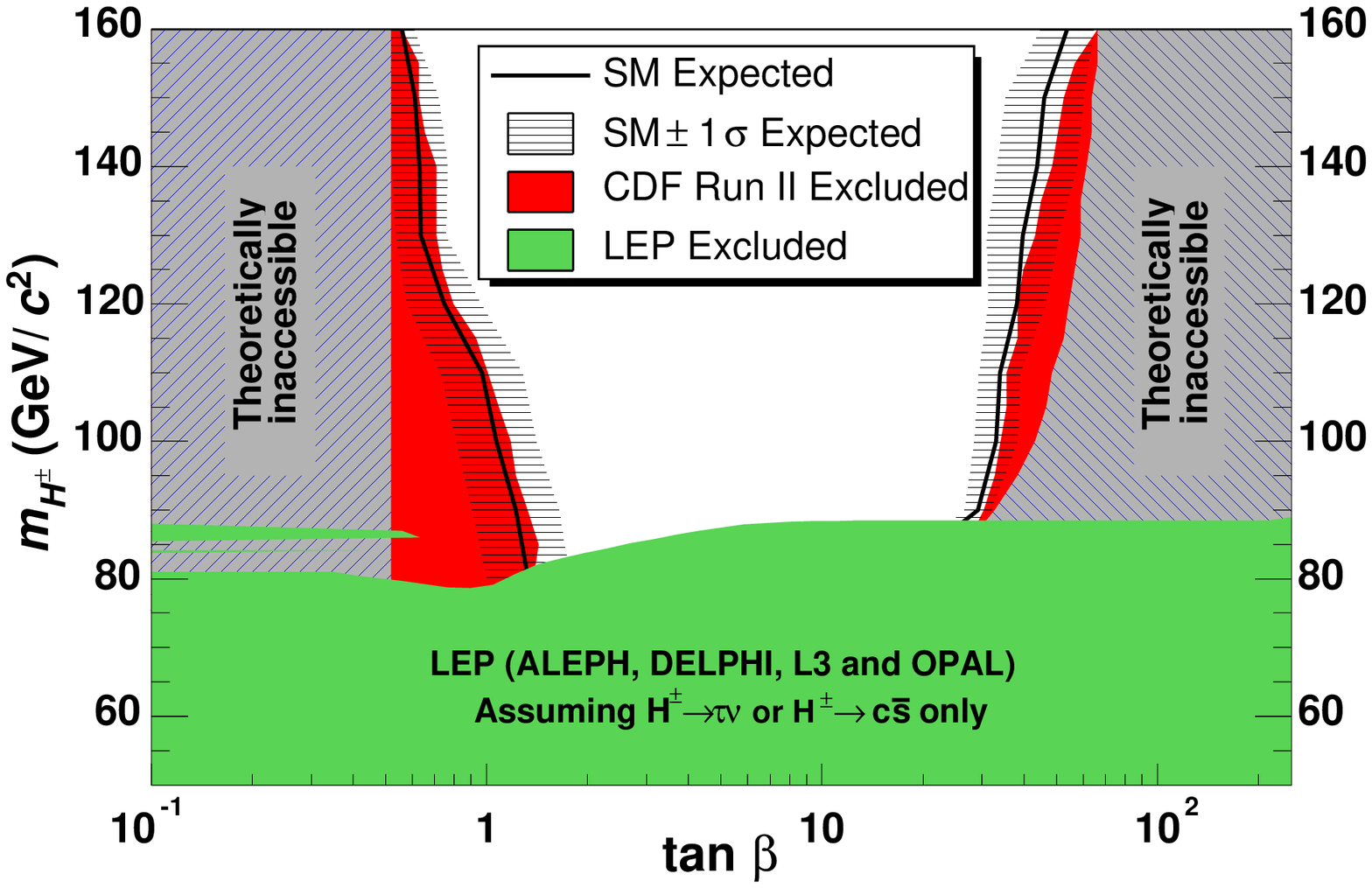}
\caption{The MSSM results obtained with 193 \invpb at CDF. The SM-expected exclusion limits are indicated by black solid lines and the $\pm 1 \sigma$ confidence band around it is obtained by generating pseudo-experiments. The darkest solid region represents the area excluded at  \ninefivecl  The solid lower region  is the LEP combined results from direct searches \cite{Eidelman:2004wy}. 
Other relevant MSSM parameters are detailed in the caption of Fig.~\ref{ExpectedEvents_fig}.\label{Res_data_B1_fig} }
\end{figure}

Using the number of events observed in the data and repeating this procedure for different Higgs masses results in the exclusion region shown in Fig.~\ref{Res_data_B1_fig}. We determine this exclusion region for several sets of benchmark parameters, including the maximal and minimal light Higgs mass scenarios described in \cite{Carena:1999xa}. The complete characterization of these scenarios and their results are described in \cite{Eusebi_thesis_cite}. In all the benchmarks used, the low \tanb region is excluded in a similar region as shown in Fig.~\ref{Res_data_B1_fig}. The high \tanb exclusion region, however, can be significantly reduced and even vanishes, due to parameters of particular benchmarks that suppress \brttoHbs. 
The obtained exclusion limits strongly depend on the prior probability used. Using a flat prior in \tanbs, which is characterized by sudden changes in the top and charged Higgs branching ratios, yields significantly different exclusion regions. It is important to note that even if all the corrections were turned off, and tree-level calculations were used, the results would be significantly stronger than those obtained in \cite{DO_PRL_cite} under the same conditions.

In the high \tanb region the decay \Htotaunu is expected to dominate in a large fraction of the MSSM parameter space. In this region the tauonic Higgs model is a good approximation, and we explicitly set BR(\Htotaunus) = 1 and evaluate the posterior probability as a function of \brttoHbs.  The value of $\Gamma_{\Hpms}$ has little effect on the results as width corrections to the efficiency are small; we set  $\Gamma_{\Hpm}=1.4$ \gevccs. The width of the top is set to  $\Gamma_{t}=\frac{1.4~\gevcc }{ {\rm 1-BR(}\ttoHbs{\rm)}}$, and the value of $m_{h^0}$ is irrelevant in this model.
We perform the scan in \brttoHb from 0 to 1.  A posterior probability density of \brttoHb is obtained using a flat prior that is constant between 0 and 1  and null elsewhere. The \ninefivecl is obtained by integrating the posterior over the maximum density region. 
This procedure is repeated for different charged Higgs masses. In the region 80 \gevcc $\leq \mH \leq 160$ \gevcc we exclude  $\brttoHbs>0.4$ at \ninefivecl 

\begin{figure}
\center
\includegraphics[width=\figsize]{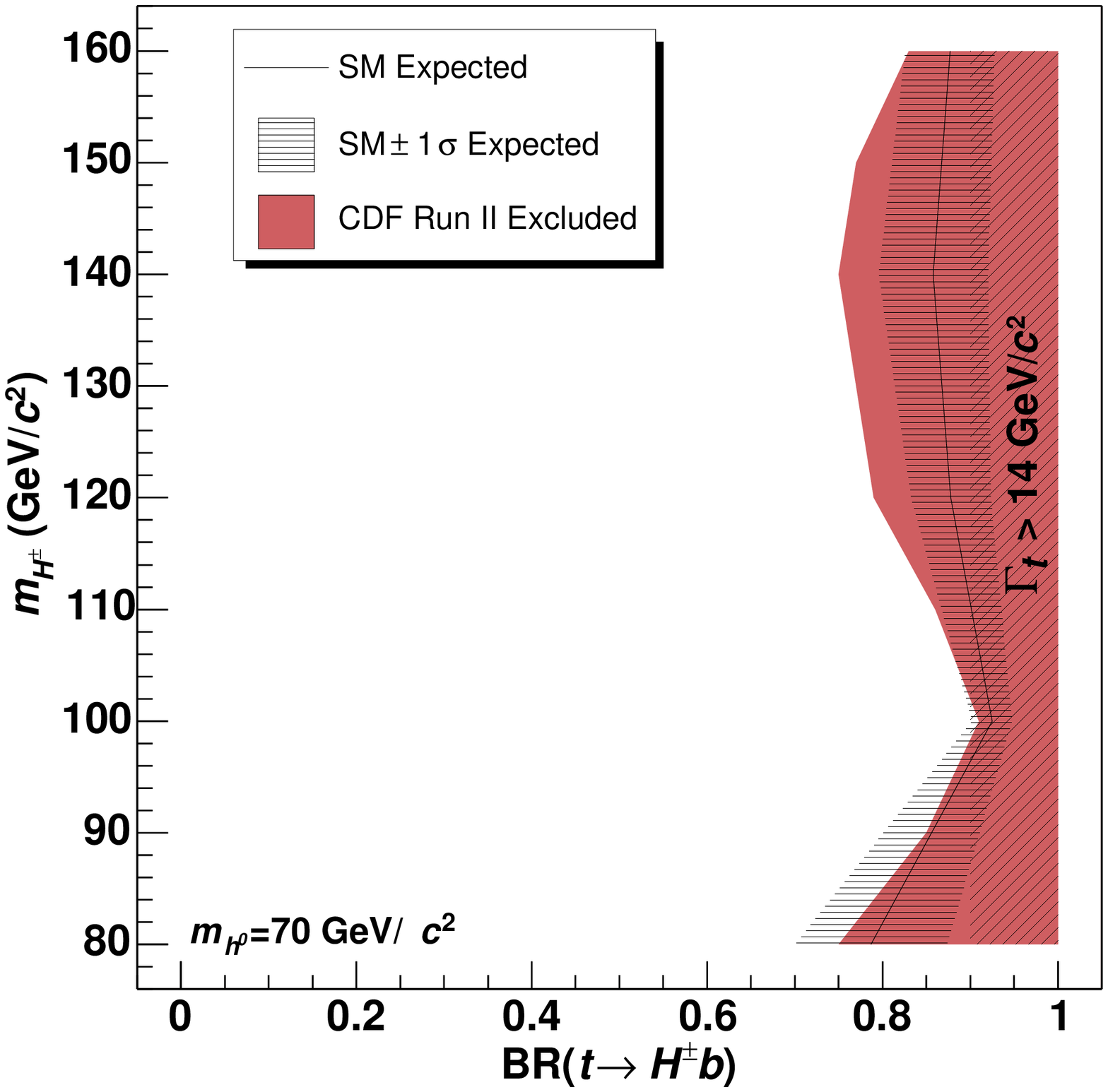}
\caption{Results for the charged Higgs branching ratio independent analysis with $\mt=175$ \gevccs. The dark solid region represents the CDF Run II excluded region in the (\mHs,\brttoHbs) plane. 
The expected exclusion limits are indicated by a black solid line and the $\pm 1 \sigma$ confidence band around it is obtained by generating pseudo-experiments. The hatched region of $\brttoHbs>0.9$ indicates that the width of the top is larger than 14 \gevccs. \label{WorstCaseScenario_fig} }
\end{figure}

Finally, in order to reduce the model dependence, we place limits on \brttoHb that hold for any combination of charged Higgs branching ratios. 
For a specific charged Higgs mass we divide each of the five charged Higgs branching ratios into 21 bins. This results in 1771 possible branching ratio combinations subject to the relation $\sum_i {\rm BR}(H^+\rightarrow X_i)=1$. For each combination, we obtain a limit on \brttoHb assuming ${\rm BR}(h^0\rightarrow b\bar{b})=0.9$ and $m_{h^0}=70$ \gevccs.  The least restrictive limit is quoted and the analysis is repeated for different charged Higgs masses. The results are shown in Fig.~\ref{WorstCaseScenario_fig}. For $\brttoHbs>0.9$ (hatched region) the width of the top is larger than 14 \gevcc and the analytical corrections to the efficiencies start losing accuracy.

In summary, we have performed a search for a charged Higgs boson in top quark decays using measurements of the top pair production cross section in four different final states and we find no evidence of signal in the region 80 \gevcc $\leq \mH \leq 160$ \gevccs. In the context of the  MSSM with full radiative corrections we exclude the low \tanb region for all benchmarks in \cite{Eusebi_thesis_cite}. The high \tanb region cannot be excluded independent of the MSSM parameters. In the tauonic Higgs model, in which the charged Higgs decays  exclusively to \taunus, the \brttoHb is  constrained to be less than 0.4 at \ninefivecl If no assumption is made on the charged Higgs decay, the \brttoHb is constrained to be less than 0.91 at \ninefivecl

\begin{acknowledgments}
We thank the Fermilab staff and the technical staffs of the participating institutions for their vital contributions. This work was supported by the U.S. Department of Energy and National Science Foundation; the Italian Istituto Nazionale di Fisica Nucleare; the Ministry of Education, Culture, Sports, Science and Technology of Japan; the Natural Sciences and Engineering Research Council of Canada; the National Science Council of the Republic of China; the Swiss National Science Foundation; the A.P. Sloan Foundation; the Bundesministerium f\"ur Bildung und Forschung, Germany; the Korean Science and Engineering Foundation and the Korean Research Foundation; the Particle Physics and Astronomy Research Council and the Royal Society, UK; the Russian Foundation for Basic Research; the Comisi\'on Interministerial de Ciencia y Tecnolog\'{\i}a, Spain; in part by the European Community's Human Potential Programme under contract HPRN-CT-2002-00292; and the Academy of Finland. 
\end{acknowledgments}

\bibliography{prl_v3.3}

\begin{thebibliography}{21}
\expandafter\ifx\csname natexlab\endcsname\relax\def\natexlab#1{#1}\fi
\expandafter\ifx\csname bibnamefont\endcsname\relax
  \def\bibnamefont#1{#1}\fi
\expandafter\ifx\csname bibfnamefont\endcsname\relax
  \def\bibfnamefont#1{#1}\fi
\expandafter\ifx\csname citenamefont\endcsname\relax
  \def\citenamefont#1{#1}\fi
\expandafter\ifx\csname url\endcsname\relax
  \def\url#1{\texttt{#1}}\fi
\expandafter\ifx\csname urlprefix\endcsname\relax\def\urlprefix{URL }\fi
\providecommand{\bibinfo}[2]{#2}
\providecommand{\eprint}[2][]{\url{#2}}

\bibitem[{\citenamefont{Higgs}(1964)}]{Higgs:1964ia}
\bibinfo{author}{\bibfnamefont{P.~W.} \bibnamefont{Higgs}},
  \bibinfo{journal}{Phys. Lett.} \textbf{\bibinfo{volume}{12}},
  \bibinfo{pages}{132} (\bibinfo{year}{1964}).

\bibitem[{\citenamefont{Nilles}(1984)}]{MSSM_cite}
\bibinfo{author}{\bibfnamefont{H.~P.} \bibnamefont{Nilles}},
  \bibinfo{journal}{Phys. Rept.} \textbf{\bibinfo{volume}{110}},
  \bibinfo{pages}{1} (\bibinfo{year}{1984}).

\bibitem[{\citenamefont{Carena and Haber}(2003)}]{HBTAP_cite}
\bibinfo{author}{\bibfnamefont{M.}~\bibnamefont{Carena}} \bibnamefont{and}
  \bibinfo{author}{\bibfnamefont{H.~E.} \bibnamefont{Haber}},
  \bibinfo{journal}{Prog. Part. Nucl. Phys.} \textbf{\bibinfo{volume}{50}},
  \bibinfo{pages}{63} (\bibinfo{year}{2003}), \eprint{hep-ph/0208209}.

\bibitem[{\citenamefont{Kidonakis and Vogt}(2003)}]{THEO_SIGMA_cite1}
\bibinfo{author}{\bibfnamefont{N.}~\bibnamefont{Kidonakis}} \bibnamefont{and}
  \bibinfo{author}{\bibfnamefont{R.}~\bibnamefont{Vogt}},
  \bibinfo{journal}{Phys. Rev.} \textbf{\bibinfo{volume}{D68}},
  \bibinfo{pages}{114014} (\bibinfo{year}{2003}), \eprint{hep-ph/0308222}.

\bibitem[{\citenamefont{Cacciari et~al.}(2004)\citenamefont{Cacciari, Frixione,
  Mangano, Nason, and Ridolfi}}]{THEO_SIGMA_cite2}
\bibinfo{author}{\bibfnamefont{M.}~\bibnamefont{Cacciari}},
  \bibinfo{author}{\bibfnamefont{S.}~\bibnamefont{Frixione}},
  \bibinfo{author}{\bibfnamefont{M.~L.} \bibnamefont{Mangano}},
  \bibinfo{author}{\bibfnamefont{P.}~\bibnamefont{Nason}}, \bibnamefont{and}
  \bibinfo{author}{\bibfnamefont{G.}~\bibnamefont{Ridolfi}},
  \bibinfo{journal}{J. High Energy Phys.} \textbf{\bibinfo{volume}{0404}},
  \bibinfo{pages}{068} (\bibinfo{year}{2004}), \eprint{hep-ph/0303085}.

\bibitem[{MET()}]{MET_cite}
\bibinfo{note}{We use a cylindrical coordinate system about the beam axis in
  which $\theta$ is the polar angle. We define transverse momentum
  $p_{T}=p\sin{\theta}$ and transverse energy $E_{T}$ similarly. Missing
  transverse energy (\vecmet) is defined as $-\sum_{i}E_{T}^{i}\hat{n}_{i}$,
  where $\hat{n}_{i}$ is the unit vector in the azimuthal plane that points
  from the $p\bar{p}$ interaction region to the $i^{th}$ calorimeter tower.
  \vecmet is further corrected for the energy and momentum of identified
  muons.}

\bibitem[{\citenamefont{Affolder et~al.}(2000)}]{CDF_PRD_cite}
\bibinfo{author}{\bibfnamefont{T.}~\bibnamefont{Affolder}} \bibnamefont{et~al.}
  (\bibinfo{collaboration}{CDF Collaboration}), \bibinfo{journal}{Phys. Rev.
  D.} \textbf{\bibinfo{volume}{62}}, \bibinfo{pages}{012004}
  (\bibinfo{year}{2000}).

\bibitem[{\citenamefont{Abe et~al.}(1997)}]{CDF_PRL_cite}
\bibinfo{author}{\bibfnamefont{F.}~\bibnamefont{Abe}} \bibnamefont{et~al.}
  (\bibinfo{collaboration}{CDF Collaboration}), \bibinfo{journal}{Phys. Rev.
  Lett.} \textbf{\bibinfo{volume}{79}}, \bibinfo{pages}{357}
  (\bibinfo{year}{1997}).

\bibitem[{\citenamefont{Abazov et~al.}(2002)}]{DO_PRL2_cite}
\bibinfo{author}{\bibfnamefont{V.~M.} \bibnamefont{Abazov}}
  \bibnamefont{et~al.} (\bibinfo{collaboration}{D0 Collaboration}),
  \bibinfo{journal}{Phys. Rev. Lett.} \textbf{\bibinfo{volume}{88}},
  \bibinfo{pages}{151803} (\bibinfo{year}{2002}), \eprint{hep-ex/0102039}.

\bibitem[{\citenamefont{Abbott et~al.}(1999)}]{DO_PRL_cite}
\bibinfo{author}{\bibfnamefont{B.}~\bibnamefont{Abbott}} \bibnamefont{et~al.}
  (\bibinfo{collaboration}{D0 Collaboration}), \bibinfo{journal}{Phys. Rev.
  Lett.} \textbf{\bibinfo{volume}{82}}, \bibinfo{pages}{4975}
  (\bibinfo{year}{1999}).

\bibitem[{\citenamefont{Carena et~al.}(2000)\citenamefont{Carena, Garcia,
  Nierste, and Wagner}}]{CARENA_cite}
\bibinfo{author}{\bibfnamefont{M.}~\bibnamefont{Carena}},
  \bibinfo{author}{\bibfnamefont{D.}~\bibnamefont{Garcia}},
  \bibinfo{author}{\bibfnamefont{U.}~\bibnamefont{Nierste}}, \bibnamefont{and}
  \bibinfo{author}{\bibfnamefont{C.~E.~M.} \bibnamefont{Wagner}},
  \bibinfo{journal}{Nucl. Phys.} \textbf{\bibinfo{volume}{B577}},
  \bibinfo{pages}{88} (\bibinfo{year}{2000}), \eprint{hep-ph/9912516}.

\bibitem[{TAG()}]{TAG_cite}
\bibinfo{note}{{A} jet is determined to be tagged if it shows a displaced
  secondary vertex. These jets typically originate from the decay of long lived
  mesons such as those resulting after the hadronization process of the b
  quark.}

\bibitem[{\citenamefont{Acosta et~al.}(2004)}]{DILEP_XS_cite}
\bibinfo{author}{\bibfnamefont{D.}~\bibnamefont{Acosta}} \bibnamefont{et~al.}
  (\bibinfo{collaboration}{CDF Collaboration}), \bibinfo{journal}{Phys. Rev.
  Lett.} \textbf{\bibinfo{volume}{93}}, \bibinfo{pages}{142001}
  (\bibinfo{year}{2004}).

\bibitem[{\citenamefont{Abulencia et~al.}(2005)}]{LTAUH_XS_cite}
\bibinfo{author}{\bibfnamefont{A.}~\bibnamefont{Abulencia}}
  \bibnamefont{et~al.} (\bibinfo{collaboration}{CDF}) (\bibinfo{year}{2005}),
  \eprint{hep-ex/0510063}.

\bibitem[{\citenamefont{Acosta et~al.}(2005)}]{LJETS1_XS_cite}
\bibinfo{author}{\bibfnamefont{D.}~\bibnamefont{Acosta}} \bibnamefont{et~al.}
  (\bibinfo{collaboration}{CDF Collaboration}), \bibinfo{journal}{Phys. Rev.}
  \textbf{\bibinfo{volume}{D71}}, \bibinfo{pages}{052003}
  (\bibinfo{year}{2005}), \eprint{hep-ex/0410041}.

\bibitem[{\citenamefont{Sjostrand et~al.}(2001)}]{PYTHIA_cite}
\bibinfo{author}{\bibfnamefont{T.}~\bibnamefont{Sjostrand}}
  \bibnamefont{et~al.}, \bibinfo{journal}{Comput. Phys. Commun.}
  \textbf{\bibinfo{volume}{135}}, \bibinfo{pages}{238} (\bibinfo{year}{2001}),
  \eprint{hep-ph/0010017}.

\bibitem[{\citenamefont{Lee et~al.}(2004)}]{CPSUPERH_cite}
\bibinfo{author}{\bibfnamefont{J.~S.} \bibnamefont{Lee}} \bibnamefont{et~al.},
  \bibinfo{journal}{Comput. Phys. Commun.} \textbf{\bibinfo{volume}{156}},
  \bibinfo{pages}{283} (\bibinfo{year}{2004}), \eprint{hep-ph/0307377}.

\bibitem[{\citenamefont{{For an introduction, see J. F. Gunion, H. Haber, G.
  Kane, and S. Dawson}}(1989)}]{Gunion:1989we}
\bibinfo{author}{\bibnamefont{{For an introduction, see J. F. Gunion, H. Haber,
  G. Kane, and S. Dawson}}}, \emph{\bibinfo{title}{The Higgs Hunter's Guide}},
  Frontiers in Physics (\bibinfo{publisher}{Addison-Wesley},
  \bibinfo{address}{Redwood City, CA}, \bibinfo{year}{1989}).

\bibitem[{\citenamefont{Eidelman et~al.}(2004)}]{Eidelman:2004wy}
\bibinfo{author}{\bibfnamefont{S.}~\bibnamefont{Eidelman}} \bibnamefont{et~al.}
  (\bibinfo{collaboration}{Particle Data Group}), \bibinfo{journal}{Phys.
  Lett.} \textbf{\bibinfo{volume}{B592}}, \bibinfo{pages}{1}
  (\bibinfo{year}{2004}).

\bibitem[{\citenamefont{Carena et~al.}(1999)\citenamefont{Carena, Heinemeyer,
  Wagner, and Weiglein}}]{Carena:1999xa}
\bibinfo{author}{\bibfnamefont{M.}~\bibnamefont{Carena}},
  \bibinfo{author}{\bibfnamefont{S.}~\bibnamefont{Heinemeyer}},
  \bibinfo{author}{\bibfnamefont{C.~E.~M.} \bibnamefont{Wagner}},
  \bibnamefont{and} \bibinfo{author}{\bibfnamefont{G.}~\bibnamefont{Weiglein}}
  (\bibinfo{year}{1999}), \eprint{hep-ph/9912223}.

\bibitem[{\citenamefont{Eusebi}(2005)}]{Eusebi_thesis_cite}
\bibinfo{author}{\bibfnamefont{R.}~\bibnamefont{Eusebi}}, Ph.D. thesis,
  \bibinfo{school}{University of Rochester} (\bibinfo{year}{2005}).

\end{thebibliography}

\end{document}